\documentclass[letterpaper,11pt]{AAS}	

\usepackage{bm}
\usepackage{amsmath}
\usepackage{amssymb}
\usepackage{subfigure}
\usepackage{float}
\usepackage[colorlinks=true, pdfstartview=FitV, linkcolor=black, citecolor= black, urlcolor= black]{hyperref}
\usepackage{overcite}
\usepackage{footnpag}			      	

\PaperNumber{XX-XXX}
\newtheorem{theorem}{Theorem}
\newtheorem{lemma}{Lemma}

\begin{document}

\title{Attitude Control of Spacecraft Formations subject to Distributed Communication Delays}

\author{Siddharth Nair\thanks{Undergraduate, Department of Aerospace Engineering, Indian Institute of Technology, Bombay, 400076. This work was performed as part of an internship at the Aerospace Systems Laboratory in UT Arlington in Summer 2016},~~
Kamesh Subbarao\thanks{AAS Senior Member, Associate Professor, Department of Mechanical \& Aerospace Engineering, The University of Texas at Arlington, 19018}
}

\maketitle{}

\begin{abstract}

This paper considers the problem of achieving attitude consensus in spacecraft formations with bounded, time-varying communication delays between spacecraft connected as specified by a  strongly connected topology. A state feedback controller is proposed and investigated using a time domain approach (via LMIs) and a frequency domain approach (via the small-gain theorem) to obtain delay dependent stability criteria to achieve the desired consensus. Simulations are presented to demonstrate the application of the strategy in a specific scenario. 

\end{abstract}

\section{Introduction}
The consensus problem of multi-agent systems has been receiving significant attention in the recent years.\cite{olfati2004 , cren2007}.  Coordination between multiple agents opens avenues for spacecraft applications such as formation control\cite{Wang1999 , VanDyke2006, mesbahi}, attitude alignment\cite{rendist}, rendezvous\cite{morse} etc.

A set of dynamical systems is said to be synchronized when there is a complete match of configuration variables describing each system. Synchronization when pertaining to a spacecraft formation refers to the state when all the spacecraft possess a common attitude. \cite{chung2009} formulates the spacecraft formation problem in a Lagrangian framework and proposes a decentralized tracking law to achieve synchronization. \cite{Wang1999} proposes a leader-follower configuration and solves the synchronization problem by designing a control strategy to make all the spacecraft track the leader(reference) spacecraft. \cite{rendist} solves the attitude synchronization problem for a wider class of communication topologies (directed graphs).

In practical applications, the flow of information between spacecraft is delayed. This is often attributed to the delay in receiving data from the other spacecraft or processing of data. \cite{Chinese2012} considers the problem of synchronization of a spacecraft formation subject to constant communication delays and solves the same by using feedback linearization to track a reference trajactory while ~\cite{Shihua} approaches the problem by proposing a backstepping controller that tracks a virtual angular velocity to achieve synchronization. \cite{nazari2016} proposes a decentralized control approach for the same problem by appealing to the geometric structure of the configuration space of the spacecraft formation. Time-varying communication delays are considered in \cite{2Chinese2012} along with delays in self-tracking control parts. This renders the dynamics nonlinear and the authors resort to constructing linear filters to derive an output feedback law.

This paper considers a spacecraft formation problem wherein there are asymmetric, bounded and time-varying delays in the communication links between the spacecraft while the feedback from the spacecraft's own states is instantaneous. To the best of our knowledge, this problem hasn't been attempted before and we propose a solution by linearizing the dynamics of the spacecraft formation system using the fact that the feedback is instantaneous. A controller is proposed for the delayed linear system to achieve consensus, followed by stability analysis of the system using both time domain and  frequency domain approaches. Simulations at the end show the results of using the proposed controller in a formation containing four spacecraft.
\section{Problem Formulation}
\subsection{Modified Rodrigues Parameters}
The multiple spacecraft attitude consensus problem is considered with the states of each spacecraft being the Modified Rodrigues Parameters (MRP) ${\bm \sigma}$ and angular velocity ${\bm \omega}$ in the body frame. The attitude kinematics and the dynamics is assumed to be identical for all spacecraft in formation and is governed by the following nonlinear ordinary differential equations: 
\begin{eqnarray}
\dot{\bm \sigma}(t)& = & {\bm P}({\bm \sigma}){\bm \omega}(t) \label{attKin} \\ 
\dot{\bm \omega}(t)& = & {\bm J}^{-1}\left(-[\tilde{\bm \omega}(t)]{\bm J} {\bm \omega}(t) + {\bm \tau} \right)
\label{attDyn}
\end{eqnarray}
where ${\bm P}({\bm \sigma}) = \dfrac{1}{4}\left(2[\tilde{\bm \sigma}(t)] + 2{\bm \sigma}(t) {\bm \sigma}^{T}(t)+(1- {\bm \sigma}^{T}(t) {\bm \sigma}(t) )\mathbb{I}_{3\times 3}\right)$, 
$[\tilde{\bm a}] = \begin{bmatrix}
0 & -a_3 & a_2\\ a_3 & 0 & -a_1\\-a_2 & a_1 & 0 
\end{bmatrix}$, 

${\bm J}$ is the symmetric moment of Inertia matrix and ${\bm \tau}$ is the control torque.

The MRP is a vector defined as $${\bm \sigma} = \hat{\bm n} \tan\left(\frac{\Phi}{4}\right)$$ where $\hat{\bm n}$ is the principal axis and $\Phi$ is the principal angle as given by Euler's rotation theorem. The MRPs yield a unique representation for the attitude of the spacecraft for all principal rotations that lie in the interval $[0,~2\pi)$. At $\Phi=2\pi$, a singularity is encountered that is typically addressed using the shadow MRP set. In this paper, it is assumed that the rotations are limited to principal rotation angles less than $2\pi$.

\subsection{Feedback Linearization}
The system represented by Eqs.~(\ref{attKin}) and (\ref{attDyn}) is rewritten into a compact second order nonlinear ode:

\begin{equation}
{\bm M}({\bm \sigma})\ddot{\bm \sigma} + {\bm C}({\bm \sigma}, \dot{\bm \sigma}) \dot{\bm \sigma} = {\bm u}
\end{equation}
\label{attSecOrder}

Let ${\bm \sigma}_{1} = {\bm \sigma}$ and ${\bm \sigma}_{2} = \dot{\bm \sigma}$. Performing the transformations ${\bm \omega} = {\bm P}({\bm \sigma})^{-1} \dot{\bm \sigma}$ and ${\bm P}({\bm \sigma})^{-1}{\bm \tau} = {\bm u}$ the dynamics can be expressed as 

\begin{align}
\dot{\bm \sigma}_{1}& = {\bm \sigma}_{2} \nonumber\\
\dot{\bm \sigma}_{2}& = {\bm M}({\bm \sigma}_{1})^{-1} \left(- {\bm C}({\bm \sigma}_{1}, {\bm \sigma}_{2}){\bm \sigma}_{2} + {\bm u} \right)
\end{align} 

where ${\bm M}({\bm \sigma}_1) = {\bm P}({\bm \sigma}_1)^{-1}{\bm J} {\bm P}({\bm \sigma}_1)$ and ${\bm C}({\bm \sigma}_1, {\bm \sigma}_2) = {\bm P}({\bm \sigma}_1)^{-1}\left({\bm J} \dot{\bm P}({\bm \sigma}_1) + \widetilde{\left[{\bm P}({\bm \sigma}_1) {\bm \sigma}_2 \right]}{\bm J} {\bm P} ({\bm \sigma}_1)\right)$. 

Using full state feedback and a straightforward feedback linearization based control law, ${\bm u} = {\bm C}({\bm \sigma}_1, {\bm \sigma}_2) {\bm \sigma}_2 + {\bm M}({\bm \sigma}_1){\bm v}$, the above dynamics are further reduced to that of a double integrator as follows.

\begin{eqnarray}
\dot{\bm \sigma}_{1}&= {\bm \sigma}_{2} \nonumber\\
\dot{\bm \sigma}_{2}&= {\bm v}
\label{eq:dblinteg}
\end{eqnarray}
\subsection{Consensus of Spacecraft Formation}
Now, consider a system of $N$ spacecraft with communication pathways among them forming a strongly connected graph, each of whom is being controlled using the control law ${\bm u}(t)$ described earlier. Thus, the attitude consensus problem for $N$ nonlinear spacecraft can be posed using the simpler structure derived earlier in Eq.~(\ref{eq:dblinteg}). 

Let $\mathbf{x} = [\mathbf{x_{1}}^T\ \mathbf{x_{2}}^T]^T$ be a vector such $\mathbf{x}_{1}$ consists of the attitudes of the $N$ spacecraft while $\mathbf{x_{2}}$ consists of their derivatives i.e. $\mathbf{x_{1}} = \left[{}^1{\bm \sigma}_1^T \cdots {}^N{\bm \sigma}_1^T\right]^T$ and $\mathbf{x_{2}} = \left[{}^1{\bm \sigma}_2^T \cdots {}^N{\bm \sigma}_2^T\right]^T$. This vector defines the state of the $N$ spacecraft system and the combined dynamics can be written as: 

\begin{align}\label{dynm1}
\mathbf{\dot{x}_{1}}(t)& = \mathbf{x_{2}}(t)\nonumber\\
\mathbf{\dot{x}_{2}}(t)& = {\mathbf{u}}(t)
\end{align}  

where the vector ${\mathbf{u}} = \left[{}^1{\bm v}^T \cdots {}^N{\bm v}^T\right]^T$ is the control input obtained by cascading the control inputs of all the spacecraft.

For the system of $N$ spacecraft described above in Eq.~(\ref{dynm1}), attitude consensus is said to be achieved when $||{}^i{\bm \sigma}_{1} - {}^j{\bm \sigma}_{1}|| \rightarrow 0$ and $||{}^i{\bm \sigma}_{2} - {}^j{\bm \sigma}_{2}|| \rightarrow 0$ as $t \rightarrow \infty$ $\forall\ i,~j, ~ i \neq j$.

In the case where the spacecraft exchange state information to their connected neighbors with no delay, the following control law drives the system to consensus \cite{ren2007}. For all the analysis from here on, we will use the compact set of equations in Eq.~(\ref{dynm1}) for the $N$ connected spacecraft.

$$\mathbf{u}(t) = -\mathbf{L} \mathbf{x_{1}}(t)-\gamma\mathbf{L}\mathbf{x_{2}}(t)$$
where $\mathbf{L} = [l_{ij}]$ is the Laplacian matrix for a given communication topology~\cite{mesbahi,ren2007} and $\gamma>0$ is a damping gain.
\subsection{Proposed Control Law}
In this paper, the consensus problem is analyzed when the communication is subject to time delays i.e. the state information of the $i^\mathrm{th}$ spacecraft is relayed to the $j^\mathrm{th}$ spacecraft after a delay $\tau_{ij}$ that satisfies
\begin{align}\label{delaydyn2}
0 ~\leq~ \tau_{ij} ~\leq~ h_{ij} \quad \& \quad |\dot{\tau}_{ij}| ~\leq~  d_{ij}
\end{align}
Clearly the time delays are different along different communication paths (spacecraft pairs) and are time-varying.

To achieve consensus, the following control law is investigated.
\begin{align}\label{clm}
\mathbf{u}(t) = -\mathbf{x_{1}}(t) - \gamma\mathbf{x_{2}}(t) - \mathop{\sum_{i=1}^{N}\sum_{j=1}^{N}}_{i\neq j}{}^{ij}\mathbf{K}\mathbf{x_1}(t-\tau_{ij}) - \mathop{\sum_{i=1}^{N}\sum_{j=1}^{N}}_{i\neq j}\gamma{}^{ij}\mathbf{K}\mathbf{x_2}(t-\tau_{ij})
\end{align}

\begin{figure}[H]
\centering
\includegraphics[scale=0.5]{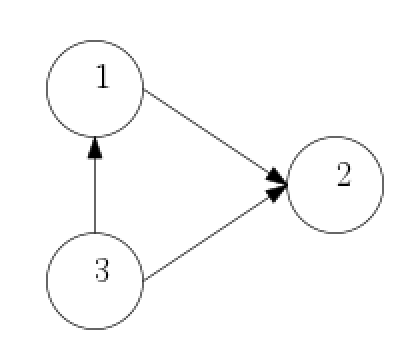}
\caption{Communication graph for the three spacecraft}
\label{fig:example}
\end{figure}

For $N$ spacecraft defining the communication topology, then for every pair $(i,~j)$ of spacecraft we have $l_{ij} \leq 0$ (from the Laplacian matrix). ${}^{ij}\mathbf{K} \in \mathbb{R}^{N \times N}$ and ${}^{ij}\mathbf{K} \otimes \mathbb{I}_{3\times 3}$  is the coefficient of the delayed state, where $\tau_{ij}$ is the time delay in sending information from the $i^\mathrm{th}$ spacecraft to the $j^\mathrm{th}$ spacecraft. The $ji^{th}$ element of ${}^{ij}\mathbf{K}$ is equal to $l_{ij}$. All other elements are zero. 
\\This is illustrated via an example involving three spacecraft with the communication topology specified by figure ~\ref{fig:example}.
The Laplacian matrix specifying the communication topology between the three spacecraft be given by $$\begin{bmatrix}1&0&-1\\-\frac{1}{2}&1&-\frac{1}{2}\\0&0&0\end{bmatrix}
$$ Let the delays in the communication channels be $\tau_{12},\ \tau_{31}$ and $\tau_{32}$. Then, the matrices $^{ij}K$ are given by $$^{12}K=\begin{bmatrix}0&0&0\\-\frac{1}{2}&0&0\\0&0&0\end{bmatrix}\otimes \mathbb{I}_{3 \times 3}\ ^{31}K=\begin{bmatrix}0&0&-1\\0&0&0\\0&0&0\end{bmatrix}\otimes \mathbb{I}_{3 \times 3}\ ^{32}K=\begin{bmatrix}0&0&0\\0&0&-\frac{1}{2}\\0&0&0\end{bmatrix}\otimes \mathbb{I}_{3 \times 3}\ $$
\\Substituting Eq.~\eqref{clm} in Eq.~\eqref{dynm1}, the state dynamics are expressed as a linear differential equation with multiple time delays.

\begin{align}\label{dynm2}
\mathbf{\dot{x}}(t) = \mathbf{A}_{0}\mathbf{x}(t) + \mathop{\sum_{i=1}^{N}\sum_{j=1}^{N}}_{i\neq j}{}^{ij}\mathbf{A}\mathbf{x}(t-\tau_{ij})
\end{align}
where \\$\mathbf{A}_{0}=\begin{bmatrix}
\mathbf{0}_{N \times N} & \mathbb{I}_{N \times N}\\-\mathbb{I}_{N \times N} & -\gamma\mathbb{I}_{N \times N}
\end{bmatrix} \otimes \mathbb{I}_{3 \times 3}$ ~and~  ${}^{ij}\mathbf{A} = \begin{bmatrix}
\mathbf{0}_{N \times N} & \mathbf{0}_{N \times N}\\
-{}^{ij}\mathbf{K} & -\gamma {}^{ij}\mathbf{K}
\end{bmatrix}\otimes \mathbb{I}_{3 \times 3}
$.\\\\
Note that $\mathop{\sum_{i=1}^{N}\sum_{j=1}^{N}}_{i\neq j}{}^{ij}\mathbf{A}= \begin{bmatrix}
\mathbf{0}_{N \times N} & \mathbf{0}_{N \times N}\\
\mathcal{A} & \gamma\mathcal{A}
\end{bmatrix}\otimes \mathbb{I}_{3 \times 3}
=\mathbf{A}_\gamma$ where $\mathcal{A}$ is the adjacency matrix of the connection topology.


\section{Stability Analysis}
\subsection{Time Domain Approach}
\begin{lemma}[Schur's Complement~\cite{gu2003}]\label{Schur}
The following linear matrix inequality (LMI) $$\mathbf{S}=\begin{bmatrix} \mathbf{S_{11}} & \mathbf{S_{12}}\\ \mathbf{S_{21}} & \mathbf{S_{22}}\end{bmatrix} < 0$$ where $\mathbf{S}= \mathbf{S}^T$, is equivalent to each of the following conditions: $$ \mathbf{S_{11}}<0,\  \mathbf{S_{22}}- \mathbf{S_{12}}^T \mathbf{S_{11}}^{-1} \mathbf{S_{12}}<0,$$ $$  \mathbf{S_{22}}<0,\  \mathbf{S_{11}}- \mathbf{S_{12}}\mathbf{S_{22}}^{-1} \mathbf{S_{12}}^{T}<0$$
\end{lemma}
\begin{lemma}[Jensen's Inequality~\cite{gu2003}]\label{Jensen}
For any constant matrix $P \in \mathbb{R}^{m\times m}$, $P=P^{T} >0 $, scalar $\gamma >0$, vector function $\omega : [0,\gamma] \rightarrow \mathbb{R}^{m}$ such that the integrations concerned are well defined, then $$\gamma\int_{0}^{\gamma}\omega^{T}(\beta)P\omega(\beta)d\beta \geq \left(\int_{0}^{\gamma}\omega(\beta)d\beta\right)^{T}P\left(\int_{0}^{\gamma}\omega(\beta)d\beta\right)$$
\end{lemma}
The lemmas stated above are key to obtaining an LMI condition for ensuring stability of the desired consensus.
\begin{theorem} 
Under the action of \eqref{clm}, the solutions to the dynamics given by \eqref{dynm1} converge to consensus if there exist symmetric, positive definite matrices  $Q_{ij} \in \mathbb{R}^{6(n-1)\times 6(n-1)}$, $S_{ij} \in \mathbb{R}^{6(n-1)\times 6(n-1)}$ and positive constant $\gamma$ such that the following LMIs hold
\begin{align*}
\Psi_1 &< \mathbf{0}\\
\Psi_2 &< \mathbf{0}\\
\Psi_3 &< \mathbf{0}
\end{align*}\\
where 
$$\Psi_1=\begin{bmatrix}
E^TEA_0+A^T_0E^TE & E^TEA_{12} &E^TEA_{13} &\ldots & E^TEA_{nn-1}\\
A^T_{12}E^TE & 0 & 0 &\ldots& 0\\
A^T_{13}E^TE & 0 & 0 & \ldots& 0\\
\vdots & \vdots & \vdots & \ddots & \vdots\\
A^T_{nn-1}E^TE & 0 & 0 &\ldots & 0
\end{bmatrix}$$
$$\Psi_2=\begin{bmatrix}
\mathop{\sum_{i=1}^{n}\sum_{j=1}^{n}}_{i\neq j}E^TQ_{ij}E & 0 &\ldots & 0 \\
0 & -E^TQ_{12}E& \vdots &\vdots\\
\vdots & 0 &\ddots & \vdots\\
0 &\ldots & & -E^TQ_{nn-1}E
\end{bmatrix}$$
$$\Psi_3=\begin{bmatrix}
\Psi_{11} & \Psi_{12} & \ldots & \Psi_{1n} & A^T_0E^T\\
\Psi_{12}^T &\ddots & \Psi_{23}\ldots &\vdots & A^T_{12}E^T\\
\vdots & \vdots & \ddots & \vdots & \vdots\\
\Psi_{1n}^T &\ldots & \ldots & \Psi_{nn-1} & A^T_{n\ n-1}E^T\\
EA_0 & EA_{12} & \ddots & EA_{n\ n-1} & -(\mathop{\sum_{i=1}^{n}\sum_{j=1}^{n}}_{i\neq j}h_{ij}S_{ij})^{-1}
\end{bmatrix}$$
where
\begin{align*}
E&=\begin{bmatrix}\left[ \mathbf{1}\ -\mathbb{I}_{n-1} \right] & \mathbf{0}\\ \mathbf{0} & \left[ \mathbf{1}\ -\mathbb{I}_{n-1} \right]\end{bmatrix}\otimes \mathbb{I}_{3 \times 3}\\
\Psi_{11}&=\mathop{\sum_{i=1}^{n}\sum_{j=1}^{n}}_{i\neq j}(\frac{1-d_{ij}}{h_{ij}}E^T(S_{ij})E)\\
\Psi_{12}&=(\frac{1-d_{12}}{h_{12}}E^T(S_{12})E)\\
\Psi_{22}&=(-\frac{1-d_{12}}{h_{12}}E^T(S_{12})E)\\
\Psi_{23}&=\mathbf{0}\\
..\textrm{and so on.}
\end{align*}
\end{theorem}
\emph{Proof}\\
Consider the following Lyapunov-Krasovskii candidate funtional with $\mathbf{y}(t)=E\mathbf{x}(t)\ (i.e., \mathbf{y}_{1}=\left[ \mathbf{1}\ -\mathbb{I}_{n-1} \right]\otimes \mathbb{I}_{3 \times 3}\ \mathbf{x}_1,\ \mathbf{y}_{2}=\left[ \mathbf{1}\ -\mathbb{I}_{n-1} \right] \otimes \mathbb{I}_{3 \times 3}\ \mathbf{x}_2$) 
\begin{align*} V=&\underbrace{\mathbf{y}(t)^{T}\mathbf{y}(t)}_{V_{1}} + \underbrace{\mathop{\sum_{i=1}^{n}\sum_{j=1}^{n}}_{i\neq j}\int_{t-\tau_{ij}}^{t}\mathbf{y}(s)^{T}Q_{ij}\mathbf{y}(s)ds}_{V_{2}}\\ &+ \underbrace{\mathop{\sum_{i=1}^{n}\sum_{j=1}^{n}}_{i\neq j}\int_{t-\tau_{ij}}^{t}\int_{\eta}^{t}\mathbf{\dot{y}}(s)^{T}S_{ij}\mathbf{\dot{y}}(s)ds d\eta}_{V_{3}} 
\end{align*}
where $Q_{ij}, S_{ij}$ are appropriately dimensioned symmetric positive definite matrices.\\
Defining $\mathbf{X}=[\mathbf{x}(t)^{T}\ \mathbf{x}^T(t-\tau_{12})\ \mathbf{x}^T(t-\tau_{13}) ...\mathbf{x}^T(t-\tau_{nn-1}) ]^{T}$, the derivatives of each individual term of the Lyapunov-Krasovskii functional are obtained as
$$\dot{V}_{1}=\mathbf{X}^{T}\Psi_1\mathbf{X}$$
$$\dot{V}_{2}=\mathbf{X}^{T}\Psi_2\mathbf{X}$$
\begin{align*}
\dot{V}_{3} &\leq \mathop{\sum_{i=1}^{n}\sum_{j=1}^{n}}_{i\neq j} h_{ij}\mathbf{\dot{x}^T}(t)E^T(S_{ij})L\mathbf{\dot{x}}(t)\\
&-\mathop{\sum_{i=1}^{n}\sum_{j=1}^{n}}_{i\neq j}(1-d_{ij})\int_{t-\tau_{ij}}^{t}\mathbf{\dot{x}^T}(s)E^T(S_{ij})L\mathbf{\dot{x}}(s)ds
\end{align*}
Using Jensen's inequality (Lemma \ref{Jensen}), the above inequality can be expressed as
\begin{align*}
\dot{V}_{3}&\leq \mathbf{X}^{T}\Omega\mathbf{X}
\end{align*}
where $$\Omega=\begin{bmatrix}
\Omega_{11} & \Omega_{12} & \ldots & \Omega_{1n}\\
\Omega_{12}^T &\ddots & \Omega_{23}\ldots &\vdots\\
\vdots & \vdots & \ddots & \vdots \\
\Omega_{1n}^T &\ldots & \ldots & \Omega_{nn-1}
\end{bmatrix}$$ and
\begin{align*}
\Omega_{11}&=\mathop{\sum_{i=1}^{n}\sum_{j=1}^{n}}_{i\neq j}(h_{ij}A^T_0E^T(S_{ij})EA_0-\frac{1-d_{ij}}{h_{ij}}E^T(S_{ij})E)\\
\Omega_{12}&=\mathop{\sum_{i=1}^{n}\sum_{j=1}^{n}}_{i\neq j}h_{ij}A^T_0E^T(S_{ij})EA_{12}+\frac{1-d_{12}}{h_{12}}E^T(S_{12})E\\
\Omega_{22}&=\mathop{\sum_{i=1}^{n}\sum_{j=1}^{n}}_{i\neq j}h_{ij}A^T_{12}E^T(S_{ij})EA_{12}-\frac{1-d_{12}}{h_{12}}E^T(S_{12})E\\
\Omega_{23}&=\mathop{\sum_{i=1}^{n}\sum_{j=1}^{n}}_{i\neq j}h_{ij}A^T_{12}E^T(S_{ij})EA_{13}\\
..\textrm{and so on.}
\end{align*}
To make the elements of the above matrix affine in the unknown variables, Schur's complement (Lemma \ref{Schur}) is applied using the fact that $\mathop{\sum_{i=1}^{n}\sum_{j=1}^{n}}_{i\neq j}h_{ij}S_{ij}$ is positive definite and invertible. The resulting LMI is
$\Psi_3=<\mathbf{0}$

The vector $\mathbf{y}(t)$ converges to $\mathbf{0}$ if there exist symmetric positive definite matrices $P,\ Q_{ij},\ S_{ij}$ such that the following LMIs are feasible
\begin{align*}
\Psi_1 &< \mathbf{0}\\
\Psi_2 &< \mathbf{0}\\
\Psi_3 &< \mathbf{0}
\end{align*}
Observe that $\mathbf{y_{1i}}(t)=\mathbf{x_{11}}(t)-\mathbf{x_{1\ i+1}}(t) \ \forall\ i >1$ . Thus, if $\mathbf{y}(t) \rightarrow \mathbf{0}$ as $t\rightarrow \infty$, we have $\mathbf{x_{11}}(t)=\mathbf{x_{1i}}(t)$ as $t\rightarrow \infty\ \forall\ i >1$. Hence, consensus is achieved.$\blacksquare$\\
Note that finding feasible solutions for the above LMIs is a tedious task considering that the LK functional is required to be positive definite on the augmented state space $\mathbf{y}(t)$ while its derivative is required to be negative definite on the actual state space $\mathbf{x}(t)$ which can't be obtained from $\mathbf{y}(t)$ via an invertible transformation due to the nature of how the time delays appear in the closed loop dynamics. To guarantee feasible solutions, a leader-follower strategy or a constant set point must be specified to obtain an invertible transformation such that the derivatives of $\mathbf{y}(t)$ appear as linear functions of $\mathbf{y}(t)$ itself, i.e, $\dot{\mathbf{Y}}(t)=\bar{\mathbf{A}}_0\mathbf{Y}(t)+\mathop{\sum_{i=1}^{N}\sum_{j=1}^{N}}_{i\neq j}{}^{ij}\bar{\mathbf{A}}\mathbf{Y}(t-\tau_{ij})$.\\
To this end, we also investigate the consensus problem using a frequency domain approach to obtain delay dependent stability criteria.

\subsection*{Frequency Domain Approach}
We adopt a frequency domain approach by making use of the small-gain theorem to show that the system achieves consensus. Before stating the main theorem, we present a few lemmas and the small-gain theorem itself.\\
Let $L^p_2$ be the space of $\mathbb{R}^p$ valued functions $f:[0,\infty)\rightarrow\mathbb{R}^p$ such that its $L_2$ norm is bounded, i.e,
$$||f||_{L_2}=\int^{\infty}_{0}f^T(t)f(t)dt<\infty$$
Let $L^p_{2e}$ be an extended space containing functions that satisfy the above inequality on finite intervals. An operator is defined as a function $G:L^p_{2e}\rightarrow L^q_{2e}$ with an induced norm defined as $$||G||=sup\left(\frac{||G(f)||}{||f||}\right): \forall f\in L^p_{2e}, f\neq0$$
\begin{figure}[H]
\centering
\includegraphics[scale=0.5]{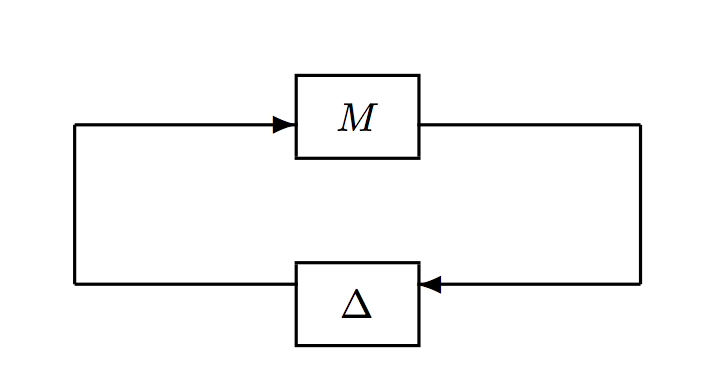}
\caption{Feedback System}
\label{fig:fb}
\end{figure}
\begin{lemma}[Small-Gain theorem] 
Suppose that $M$ and $\Delta$ are both stable systems with finite input-output gains. Let $||M||$ and $||\Delta||$ denote their respective induced norms. Then the feedback system in figure ~\ref{fig:fb} is stable if $$||M\Delta ||<1$$ 
\end{lemma}
The small-gain theorem provides a sufficient condition for stability of feedback systems. Let us examine the form of ~\eqref{dynm2} more closely.
\begin{align*}
\mathbf{\dot{x}}(t)&= \mathbf{A}_{0}\mathbf{x}(t) + \mathop{\sum_{i=1}^{N}\sum_{j=1}^{N}}_{i\neq j}{}^{ij}\mathbf{A}\mathbf{x}(t-\tau_{ij})\\
\Rightarrow\mathbf{\dot{x}}(t)&= \mathbf{A}_{0}\mathbf{x}(t) + \mathop{\sum_{i=1}^{N}\sum_{j=1}^{N}}_{i\neq j}{}^{ij}\mathbf{A}(\mathbf{x}(t-\tau_{ij})-\mathbf{x}(t))+\mathop{\sum_{i=1}^{N}\sum_{j=1}^{N}}_{i\neq j}{}^{ij}\mathbf{A}\mathbf{x}(t)\\
\Rightarrow\mathbf{\dot{x}}(t)&= \mathbf{A}_{0}\mathbf{x}(t) + \mathop{\sum_{i=1}^{N}\sum_{j=1}^{N}}_{i\neq j}{}^{ij}\mathbf{A}(\mathbf{x}(t-\tau_{ij})-\mathbf{x}(t))+\mathbf{A}_{\gamma}\mathbf{x}(t)\\
\Rightarrow\mathbf{\dot{x}}(t)-\mathbf{A}_{\gamma}\mathbf{x}(t)-\mathbf{A}_{0}\mathbf{x}(t)&= \mathop{\sum_{i=1}^{N}\sum_{j=1}^{N}}_{i\neq j}{}^{ij}\mathbf{A}(\mathbf{x}(t-\tau_{ij})-\mathbf{x}(t))\\
\end{align*}
Thus, the system of spacecraft in our problem ~\eqref{dynm2} can be viewed as a feedback system with the same structure in Figure ~\ref{fig:fb} with the  transfer function $$\mathbf{X}(s)=\mathbf{T}(s)\Delta(\mathbf{X}(s))$$ where $\mathbf{T}(s)=(s\mathbb{I}_{6N \times 6N}-\mathbf{A}_0-\mathbf{A}_\gamma)^{-1}\left[\mathbf{A}_{12}\ \mathbf{A}_{13}...\ \mathbf{A}_{nn-1}\right]$ and $\Delta=\begin{bmatrix} \Delta_{\tau_{12}}& 0 & ...& 0\\
0& \Delta_{\tau_{13}} & ... & 0\\
0&..& 0 & \Delta_{\tau_{nn-1}}\end{bmatrix}$ with delay operators $\Delta_{\tau}$ defined as $\Delta_\tau(\mathbf{x}(t))=\mathbf{x}(t)-\mathbf{x}(t-\tau)$

\begin{lemma}[~\cite{kao2007st}]\label{delayb}
Let $\mathbb{S}$ be a set of differentiable functions such that $\mathbb{S}=\left\{ \tau(t)|\tau(t)\in[0,\tau_0], \dot{\tau}(t)\leq d\ \forall t\ \right\}$. Then the delay operator obeys the following equality $$\mathop{sup}_{\tau(t)\in\mathbb{S}}(||\Delta_\tau\frac{1}{s}||)=\tau_0$$
\end{lemma}
\emph{Proof}\\
Let $v(t) \in L^p_{2e}$. Then $$\Delta_\tau\frac{1}{s} v(t)=\int^{t}_{t-\tau(t)}v(t)dt$$
Now,
$$||\Delta_\tau\frac{1}{s} v(t)||^2=(\int^{t}_{t-\tau(t)}v(s)ds)^T(\int^{t}_{t-\tau(t)}v(s)ds)$$
Using lemma ~\ref{Jensen}, we have
$$||\Delta_\tau\frac{1}{s} v(t)||^2\leq\tau_0(\int^{t}_{t-\tau_0}v^T(s)v(s)ds)$$
The $L_2$ norm of $\Delta_\tau\frac{1}{s} v(t)$ is bounded above as follows
\begin{align*}
||\Delta_\tau\frac{1}{s} v(t)||_{L_2}^2&\leq\int^{\infty}_{0}\tau_0(\int^{t}_{t-\tau_0}v^T(s)v(s)ds)dt\\
&=\tau_0\int^{0}_{-\tau_0}\int^{\infty}_{0}v^T(s)v(s)dtds\\
&\leq \tau^2_0||v||_{L_2}
\end{align*}
$\blacksquare$\\
The above lemma establishes a bound for the delay operator. The small-gain theorem also requires the plant $\mathbf{T}(s)$ to be stable. The following lemma characterizes the values of the gain $\gamma$ that render the system stable in the case where the delays are non-existent.
\begin{lemma}[~\cite{ren2007}]\label{sysb}
The system described by ~\eqref{dynm2} in the case where the delays are non-existent is stable for the values of $\gamma$ which satisfy
$$\gamma > \mathop{max}_{\mu_i\neq0}\sqrt{\frac{2}{|\mu_i|\cos(\frac{\pi}{2}-tan^{-1}(\frac{-Re\ \mu_i}{Im\ \mu_i}))}}$$
if the connection topology of the spacecraft formation has a rooted directed spanning tree.
\end{lemma}
Now we make use of the above lemmas to prove the following theorem which establishes the condition under which consensus is guaranteed.
\begin{theorem}\label{db}
The system of spacecraft with dynamics dictated by ~\eqref{dynm2} connected via a network topology containing a rooted directed spanning tree, achieves consensus if along with the conditions of lemmas ~\ref{delayb} and ~\ref{sysb}, the following holds
$$\tau_0 < \frac{1}{\omega\ \mathop{max}_{\mu_i}\sum^{p_i}_{k=1}|(j\omega-\frac{\gamma\mu_i\pm\sqrt{\gamma^2\mu^2_i+4\mu_i}}{2})^{-k}|(1+\gamma)}\quad \forall\omega\in(0,\infty)$$
where $\mu_i$ are the eigenvalues of $-\mathbf{L}\otimes \mathbb{I}_{3 \times 3}$, $p_i$ are the multiplicities of the corresponding eigenvalues of $\mathbf{A}_0+\mathbf{A}_{\gamma}$ and $\tau_0=\mathop{max}_{i,j}h_{ij}$.
\end{theorem}
\emph{Proof}\\
Now that we have that the individual transfer functions $\mathbf{T}(s)$ and $\Delta$ are stable with finite input-output gains, we proceed to find a bound on $||\mathbf{T}\Delta||$ and subsequently use the small-gain theorem to show that consensus is achieved for all signals with frequencies $\omega\in [0,\infty)$
\begin{align*}
||T(s)\Delta||&=||(s\mathbb{I}_{6N \times 6N}-\mathbf{A}_0-\mathbf{A}_\gamma)^{-1}\left[s\mathbf{A}_{12}\ s\mathbf{A}_{13}...\ s\mathbf{A}_{nn-1}\right]\Delta\frac{1}{s}||\\
&=||(s\mathbb{I}_{6N \times 6N}-\mathbf{A}_0-\mathbf{A}_\gamma)^{-1}(\mathop{\sum_{i=1}^{N}\sum_{j=1}^{N}}_{i\neq j}s\mathbf{A}_{ij}\Delta_{\tau_{ij}}\frac{1}{s})||\\
&\leq \mathop{sup}_{\omega}\left(||(s\mathbb{I}_{6N \times 6N}-\mathbf{A}_0-\mathbf{A}_\gamma)^{-1}s\mathbf{A}_{\gamma}||\right)\times\mathop{max}_{i,j}h_{ij}
\end{align*}
Since the sum of elements in each row of the adjacency matrix $\mathcal{A}$ is 1, we have $||\mathbf{A}_{\gamma}||_{\infty}=1+\gamma$. Also observe that det$(s\mathbb{I}_{6N \times 6N}-\mathbf{A}_0-\mathbf{A}_\gamma)=$det$(s^2\mathbb{I}_{3N \times 3N}+s\gamma\mathbf{L}\otimes \mathbb{I}_{3 \times 3}+\mathbf{L}\otimes \mathbb{I}_{3 \times 3})$. Thus the eigenvalues of $\mathbf{A}_0+\mathbf{A}_{\gamma}$ are given by $\lambda_i=\frac{\gamma\mu_i\pm\sqrt{\gamma^2\mu^2_i+4\mu_i}}{2}$ where $\mu_i$ are the eigenvalues of $-\mathbf{L}\otimes \mathbb{I}_{3 \times 3}$. \\Then, we have $||((s\mathbb{I}_{6N \times 6N}-\mathbf{A}_0-\mathbf{A}_\gamma)^{-1}||_{\infty} = \mathop{max}_{i}\sum^{p_i}_{k=1}|(s-\lambda_i)^{-k}|$~\cite{yang2008} where $p_i$ is the multiplicity of eigenvalue $\lambda_i$. Thus, to ensure that all signals with frequencies $\omega \in (0,\infty)$ die out, we enforce the small-gain theorem to get
\begin{align*}
||T(s)\Delta||&\leq\mathop{max}_{\mu_i}\sum^{p_i}_{k=1}|(j\omega-\frac{\gamma\mu_i\pm\sqrt{\gamma^2\mu^2_i+4\mu_i}}{2})^{-k}|\omega(1+\gamma)\tau_0 <1\\
\Rightarrow\tau_0& < \frac{1}{\omega\ \mathop{max}_{\mu_i}\sum^{p_i}_{k=1}|(j\omega-\frac{\gamma\mu_i\pm\sqrt{\gamma^2\mu^2_i+4\mu_i}}{2})^{-k}|(1+\gamma)}\quad \forall\omega\in(0,\infty)
\end{align*}
Under this condition, all signals with non-zero frequencies die out.
To see what happens with signals with zero frequency, i.e., $s=j\omega=0$, we take the Laplace transform of ~\eqref{dynm2} to get 
\begin{align*}
s^2\mathbf{X}(s)&=-\mathbf{X}(s)-\gamma s\mathbf{X}(s)+\mathcal{A}\otimes \mathbb{I}_{3 \times 3}\mathbf{X}(s)+\gamma\mathcal{A}\otimes \mathbb{I}_{3 \times 3}s\mathbf{X}(s) \quad  for\ s=j\omega=0\\
\Rightarrow \mathbf{L}\otimes \mathbb{I}_{3 \times 3}\mathbf{X}&=0\quad \textrm{because}\ \mathbf{L}\otimes \mathbb{I}_{3 \times 3}=\mathbb{I}_{3N \times 3N}-\mathcal{A}\otimes \mathbb{I}_{3 \times 3}
\end{align*}
Since the connection topology has a rooted directed spanning tree~\cite{cren2007}, $0$ is a simple eigenvalue of $\mathbf{L}$ with eigenvector $[1\ 1\ 1..\ 1]^T$. Thus, after the non-zero frequencies have died out, we are left with $\mathbf{X}=[c\ c\ c..\ c]^T$ for some $c\in \mathbb{R}$ and by definition, the system has achieved consensus.

$\blacksquare$

\section{Simulations}
\begin{figure}[H]
\centering
\includegraphics[scale=0.5]{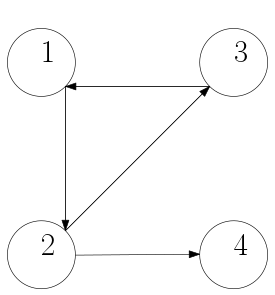}
\caption{Communication Graph}
\label{fig:graph}
\end{figure}

The proposed control strategy is implemented for a formation of $4$ spacecraft communicating with each other via the graph shown in Figure ~\ref{fig:graph}. 
The simulation conditions are as specified in the tables below.
\begin{table}[H]
\begin{tabular}{ | c | c | c | c | c |}
\hline
 & Spacecraft 1 & Spacecraft 2 & Spacecraft 3 & Spacecraft 4\\ \hline
 Inertia Matrix & $20\times\mathbb{I}_{3\times 3}$ & $30\times\mathbb{I}_{3\times 3}$ & $40\times\mathbb{I}_{3\times 3}$ & $50\times\mathbb{I}_{3\times 3}$\\ \hline
Initial attitude (MRP) &  $\begin{bmatrix}0.8\\0.8\\0.8\end{bmatrix}$ & $\begin{bmatrix}0.4\\0.4\\0.4\end{bmatrix}$ & $\begin{bmatrix}-0.6\\-0.6\\-0.6\end{bmatrix}$ & $\begin{bmatrix}-0.8\\-0.8\\-0.8\end{bmatrix}$\\ \hline
Initial Angular Velocity &  $\begin{bmatrix}0.06849\\0.06849\\0.06849\end{bmatrix}$ & $\begin{bmatrix}0\\0\\0\end{bmatrix}$ & $\begin{bmatrix}-0.09615\\-0.09615\\-0.09615\end{bmatrix}$ & $\begin{bmatrix}0.06849\\0.06849\\0.06849\end{bmatrix}$\\ \hline
\end{tabular}
\caption{ Spacecraft Initial conditions and Parameters}
\end{table}
\begin{table}[H]
\begin{tabular}{ | l | c | c | c | c |}
\hline
 & $1 \rightarrow 2$ & $2 \rightarrow 3$ & $3 \rightarrow 1$ & $2 \rightarrow 4$\\ \hline
 Bound on delay (h) & $5$ seconds & $6$ seconds & $7$ seconds & $5$ seconds\\ \hline
Bound on delay derivative (d)&  $1$ & $2$ & $0.5$ & $1$\\ \hline
\end{tabular}
\caption{ Time Delay Parameters}
\end{table}
For the given communication topology, lemma ~\ref{sysb} yields $\gamma > 1.414$. Thus, the damping gain $\gamma$ is set to $5$. Theorem ~\ref{db} yields a conservative upper bound on the delay at 9.6346 seconds. The simulation results are presented below. All vectors are expressed in body frames of the respective spacecraft.
\begin{figure}[H]
\centering
\includegraphics[scale=0.3]{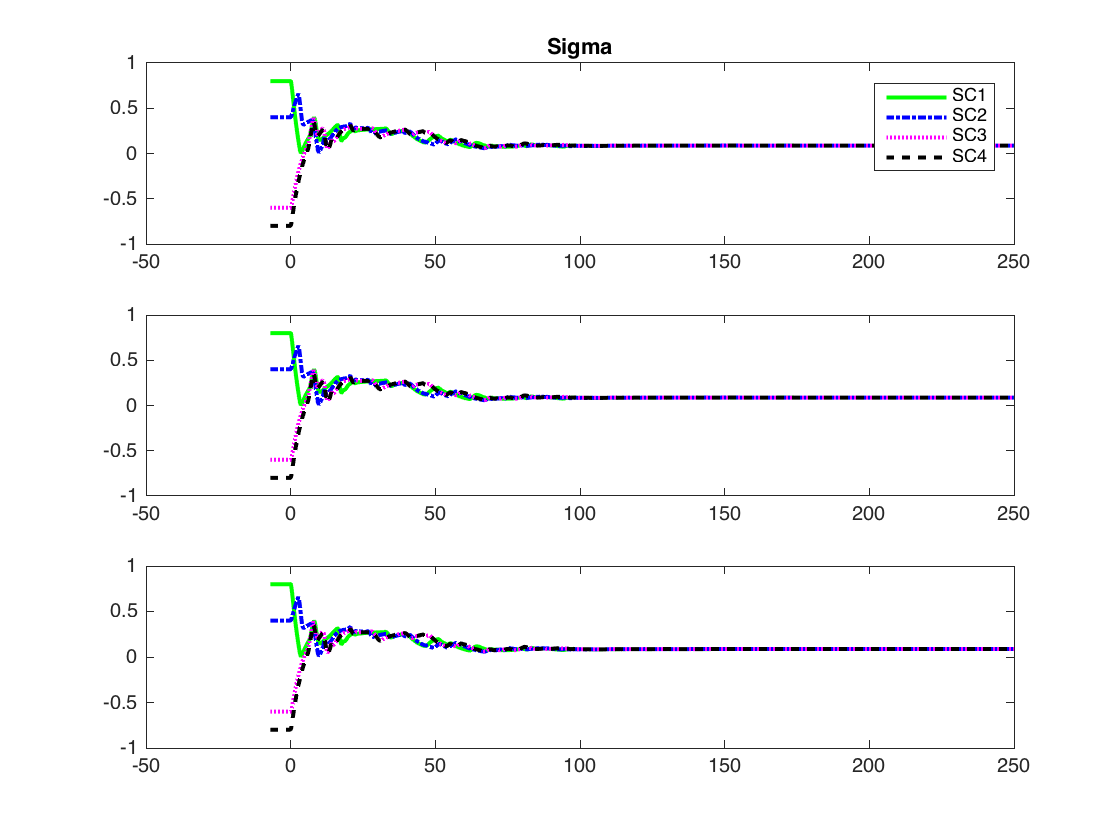}
\caption{Spacecraft Attitudes vs Time}
\end{figure}
\begin{figure}[H]
\centering
\includegraphics[scale=0.3]{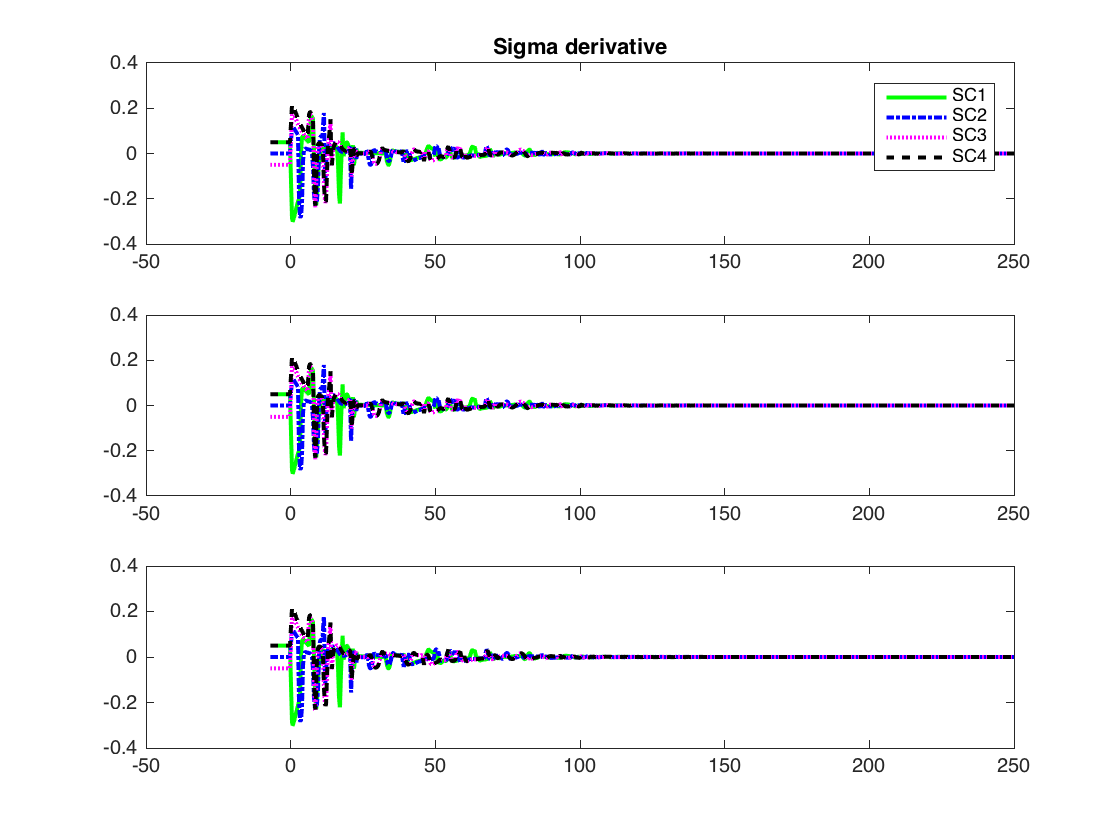}
\caption{Spacecraft Attitude Derivatives vs Time}
\end{figure}
As we can see, the spacecraft achieve consensus.
\begin{figure}[H]
\centering
\includegraphics[scale=0.3]{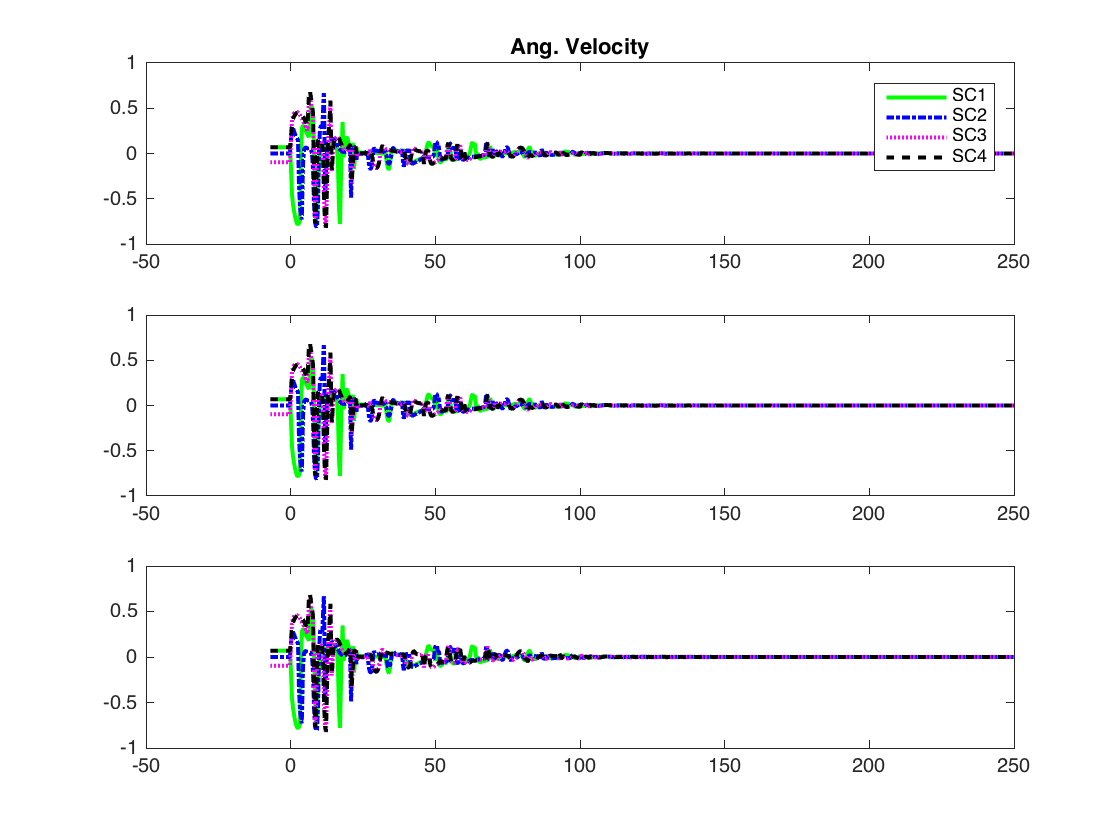}
\caption{Spacecraft Angular Velocities vs Time}
\end{figure}
\begin{figure}[H]
\centering
\includegraphics[scale=0.3]{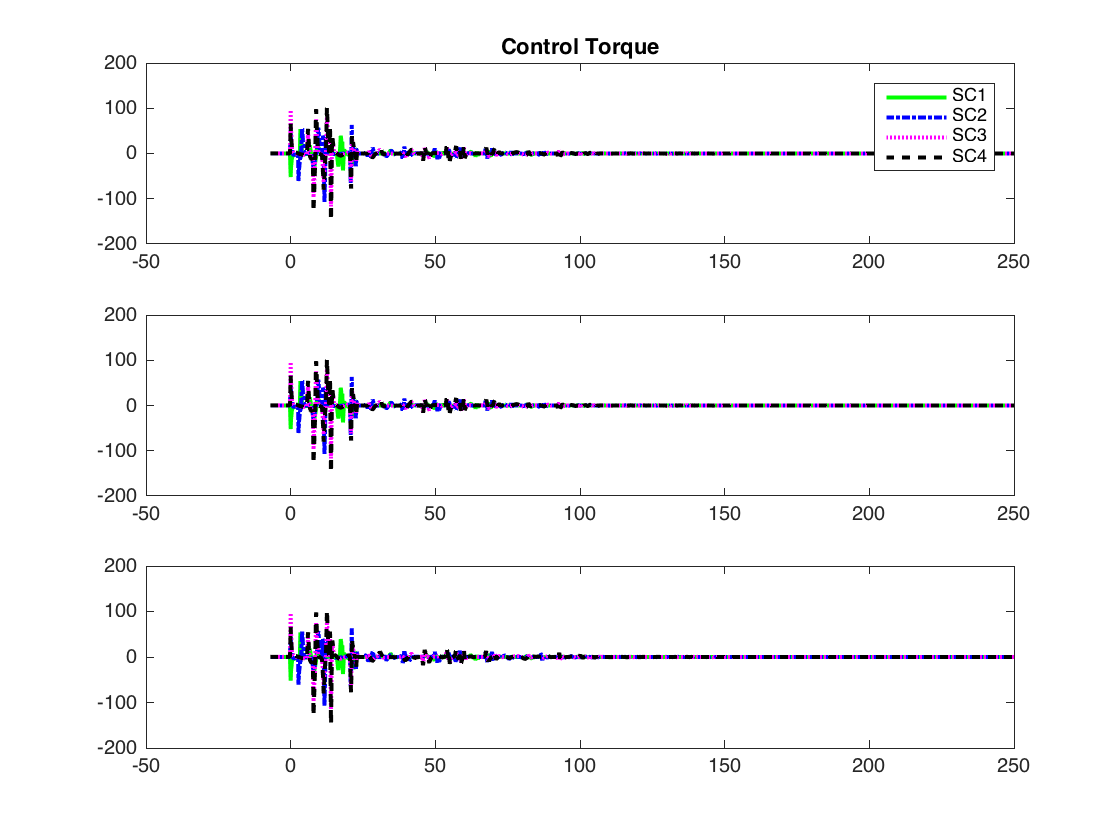}
\caption{Spacecraft Control Torques vs Time}
\end{figure}
The control torques are obtained by transforming the control input \eqref{clm} from the $\left[ {\bm \sigma}~ \dot{\bm \sigma} \right]$ space back to the $\left[ {\bm \sigma}~ {\bm \omega} \right]$ space. These are the actual torques that are implemented by the respective spacecraft towards achieving the desired consensus.
\\On setting the gain $\gamma$ to 0.1, we observe that consensus is not achieved and the states of the spacecraft diverge from each other. The results are shown below.
\begin{figure}[H]
\centering
\includegraphics[scale=0.3]{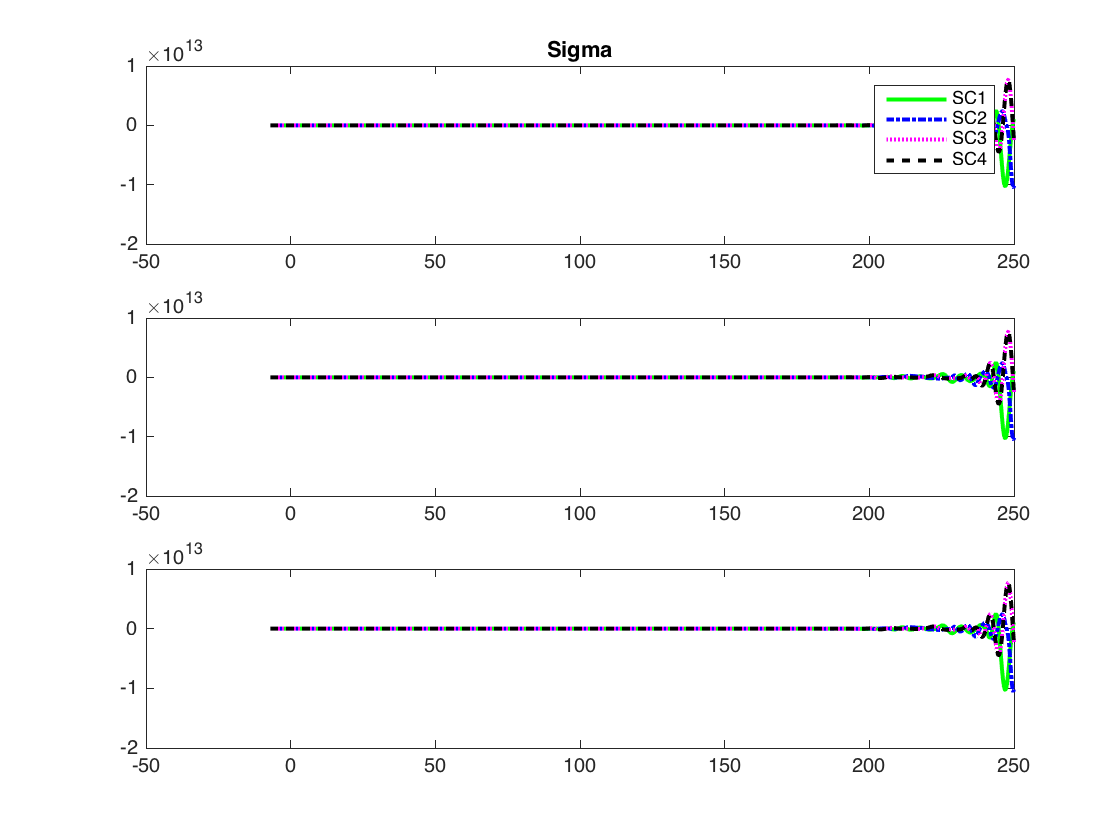}
\caption{Spacecraft Attitudes vs Time when gain $\gamma$ is set to 0.1}
\end{figure}
\begin{figure}[H]
\centering
\includegraphics[scale=0.3]{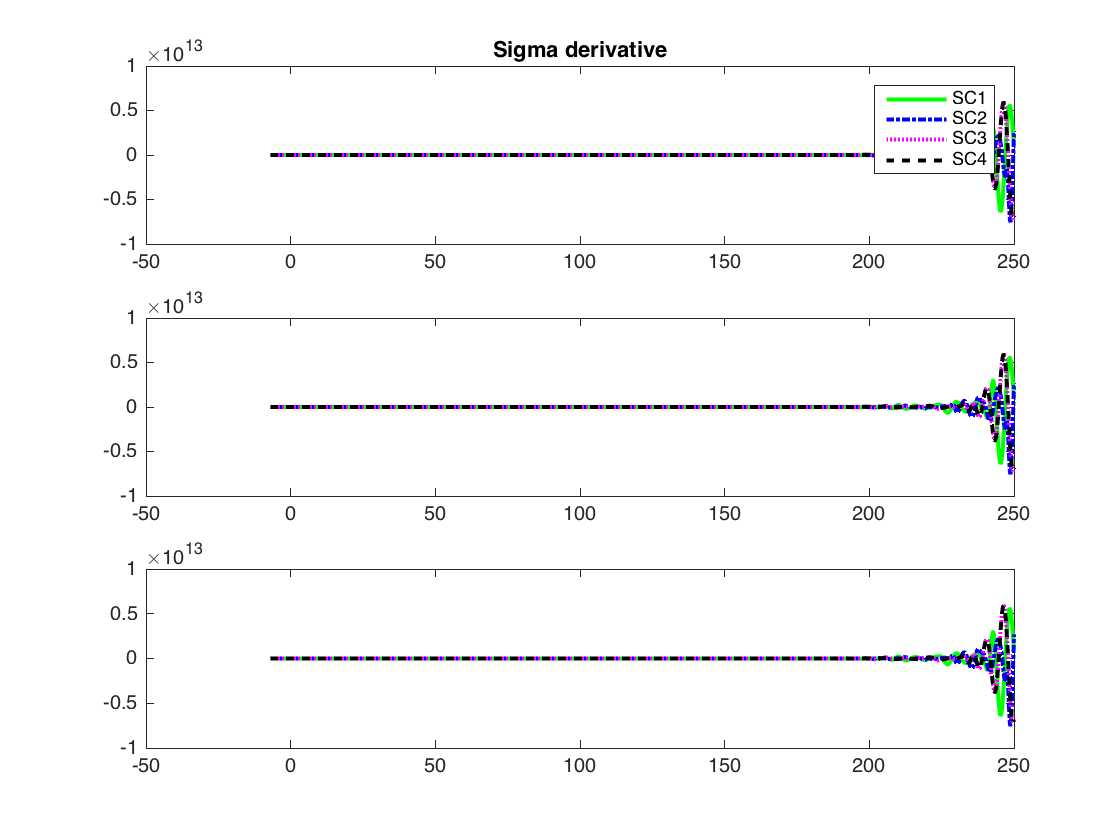}
\caption{Spacecraft Attitude Derivatives vs Time when gain $\gamma$ is set to 0.1}
\end{figure}

\section{Summary and Conclusions}
Distributed variable time-delays were considered in attitude consensus of multiple cooperating spacecraft. The mathematical proof for consensus was shown using the small-gain theorem for spacecraft formations with a rooted directed spanning tree. Simulation results for a 4 spacecraft attitude consensus scenario were presented. 
\bibliographystyle{AAS_publication}   
\bibliography{references}   

\end{document}